\documentclass[aps,reprint, prl, superscriptaddress, longbibliography,showpacs,nobibnotes,floatfix]{revtex4-1}
\usepackage{graphicx}
\usepackage{bm,amsmath}
\usepackage{dcolumn}
\usepackage{natbib,xcolor}
\usepackage[version=4]{mhchem}
\usepackage{upgreek}

\begin{document}
\title{Metamagnetism and anomalous magnetotransport properties in rare-earth-based polar semimetals $\textbf{\textit{R}}$AuGe ($\textbf{\textit{R}}$ = Dy, Ho, and Gd)}

\author{Takashi Kurumaji}
\affiliation{Division of Physics, Mathematics and Astronomy, California Institute of Technology, Pasadena, California 91125, USA}
\affiliation{Department of Advanced Materials Science, University of Tokyo, Kashiwa 277-8561, Japan}
\author{Masaki Gen}
\affiliation{RIKEN Center for Emergent Matter Science (CEMS), Wako 351-0198, Japan}
\author{Shunsuke Kitou}
\affiliation{Department of Advanced Materials Science, University of Tokyo, Kashiwa 277-8561, Japan}
\author{Taka-hisa Arima}
\affiliation{Department of Advanced Materials Science, University of Tokyo, Kashiwa 277-8561, Japan}
\affiliation{RIKEN Center for Emergent Matter Science (CEMS), Wako 351-0198, Japan}
\date{\today}
\begin{abstract}
We report the magnetic, magnetoelastic, and magnetotransport properties of single crystals of polar magnets $R$AuGe ($R=$ Dy, Ho, and Gd), grown by Au-Ge self-flux.
Magnetization and magnetostriction measurements reveal multi-step metamagnetic transitions for the $c$-axis magnetic field ($H\parallel c$) for DyAuGe and HoAuGe, suggesting magnetic frustration in the triangular lattice of $R$ ions.
The magnetic phase diagrams have clarified a close connection between the magnetoelastic property and the emergence of the intermediate metamagentic phase.
The magnetic-field dependence of the resistivity and Hall resistivity reveal the semimetallic transport dominated by hole-type carriers, consistent with the behavior in a nonmagnetic analogue YAuGe.
We also identify a signature of an anomalous Hall effect (AHE) proportional to the field-induced magnetization in $R=$ Dy, Ho, and Gd.
GdAuGe shows magnetic and transport behavior as reported in a previous study using Bi-flux grown single crystals, while the self-flux grown crystal shows larger magnetoresistance ($\sim$ 345\%, at 1.8 K and 9 T) due to higher hole-type carrier mobility ($\sim$ 6400 cm$^2$/Vs).
Using the two-band model analysis considering the mobility change during the magnetization process, we extract the anomalous Hall conductivity: $\sim 1200$ S/cm and $\sim 530$ S/cm for $R=$ Dy and Ho, respectively, at 1.8 K with 9 T for $H\parallel c$.
The magnitude of conductivity suggests a contribution of intrinsic origin, possibly related to the Berry curvature in the electron bands induced by the time-reversal symmetry breaking and the polar lattice.
\end{abstract}

\maketitle

\section{Introduction}
Equiatomic ternary $RTX$ rare-earth intermetallics, where $R$ represents rare earths, $T$ are 3d/4d transition metals, and $X$ are the p-block elements, are a group of materials that show intriguing coupling between rare-earth magnetism and electronic transport properties \cite{gupta2015review}.
A variety of crystal structures are available as a playground of topological band crossings \cite{lin2010half,chadov2010tunable,zhang2011topological,narayan2015class}, and among them, polar semimetals, which break the spatial-inversion symmetry by their crystal lattice, receive an intense interest as candidates of topological semimetals \cite{gibson2015three,cao2017dirac,gao2018dirac,gao2021noncentrosymmetric,xu2017discovery} and as a source of anomalous electrical transport through magnetic correlation \cite{du2015dirac,mondal2019broken,gaudet2021weyl}, which also potentially induces unconventional spin textures \cite{srivastava2020observation,yao2023large}.

\begin{figure}[t]
	\includegraphics[width =  \columnwidth]{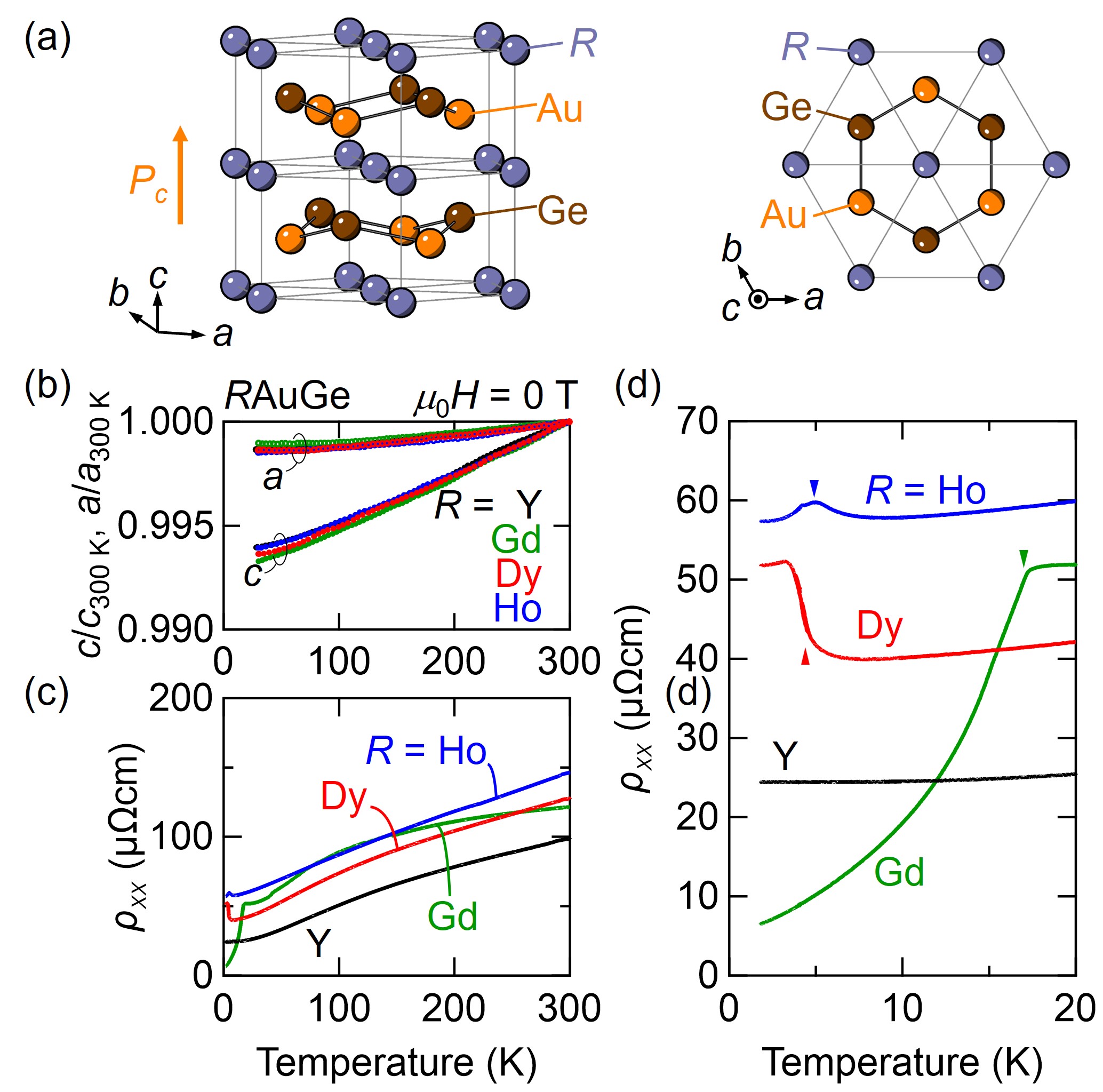}
	\caption{\label{figreshall}(a) Crystal structure of $\it{R}$AuGe ($\it{R}$ = Gd, Dy, Ho, and Y): (left) side view, (right) top view.
 The arrangement of Au and Ge breaks the inversion symmetry giving crystallographic polarity ($P_c$) along the $c$ axis.
 (b)-(c) Temperature dependence of (b) lattice constants ($a$ and $c$) normalized at $T=300$ K and (c) in-plane resistivities ($\rho_{xx}$) at zero field.
 (d) Magnified plot of (c) at low temperatures.
 Solid triangle denotes the transition temperature ($T_{\text{N1}}$) from paramagnetic state to a magnetically ordered state, determined in the magnetization measurement in Ref. \cite{kurumaji2023single}: $T_{\text{N}1}=$ 4.4 K, 4.9 K, and 17 K for $R=$ Dy, Ho, and Gd, respectively.
 }
\end{figure}

$R$AuGe ($R=$ Sc, La-Nd, Sm, Gd-Tm, and Lu) is a family of the $RTX$ intermetallic phases belonging to the polar LiGaGe-type (or isopointal NdPtSb-type) structure (space group: $P6_{3}mc$) as shown in Fig. \ref{figreshall}(a) \cite{rossi1992ternary,pottgen1996crystal}.
The nominal charge-balanced valence configuration $R^{3+}$Au$^{+}$Ge$^{4-}$ supports a pseudo-gap in the density of states of conduction electrons and semimetallic nature with compensated electron and hole bands \cite{pottgen1996crystal,schnelle1997crystal,wang2020chemical}.
The interplay between a polar nature and high electrical conductivity in epitaxially grown thin films has attracted recent interest \cite{du2019high,du2022controlling}.
In previous band calculations of nonmagnetic isostructural $R$AuGe ($R=$ Sc, Y, La, Lu) \cite{pottgen1996crystal,schnelle1997crystal}, dispersive hole bands around the $\Gamma$ point in the Brillouin zone and relatively flat electron bands along the $\Gamma-M$ and $\Gamma-K$ lines are suggested.

Using polycrystals, magnetic properties have been studied \cite{penc1999magnetic}, and resistivity and zero-field thermoelectricity have been reported \cite{gibson1996susceptibility}.
CeAuGe and TmAuGe are known to exhibit ferromagnetism \cite{pottgen1996ferromagnetic,gibson2000magnetic,kaczorowski2014magnetic}, while the other $R$AuGe ($R=$ Nd, Gd-Er) exhibit complex antiferromagnetic transitions \cite{penc1999magnetic}.
Powder neutron diffraction studies revealed that they host magnetic modulation in the $ab$-plane \cite{baran2000magnetic,baran2001neutron,gibson2001crystal}, suggesting the inherent magnetic frustration in a triangular lattice of $R$ ions (Fig. \ref{figreshall}(a)). 

Single crystal growth has been reported by our group in Ref. \cite{kurumaji2023single}, where Au-Ge self-flux was used.
The systematic evolution of the polar crystal structure among heavy $R$ compounds and the anisotropic magnetic properties were elucidated.
Another single crystal study in \ce{GdAuGe} is reported by using crystals grown by Bi-flux, and large anomalous Hall effect (AHE) was observed \cite{ram2023multiple}.

In our resonant x-ray diffraction study using single crystals \cite{kurumaji2024canted}, DyAuGe has been revealed to show an antiferromagnetic transition at $T_{\text{N1}}=4.4$ K.
The ground state at zero field is a canted antiferromagnetic state driven by the antiferroquadrupole ordering of the 4f electrons in Dy ions.
HoAuGe is also known to exhibit similar magnetism with a transition temperature at $T_{\text{N1}}=4.9$ K \cite{gibson2001crystal}, while the zero-field ground state is assigned as a conventional collinear antiferromagnetic state \cite{gibson2001crystal,kurumaji2024canted}.
GdAuGe has been reported to show an antiferromagnetic transition at $T_{\text{N1}}=17$ K \cite{pottgen1996crystal,gibson1996susceptibility}, and successive transitions and metamagnetism associated with spin-flop-like transition have been reported for $H\parallel c$ \cite{ram2023multiple}.

Here, we study magnetic, magnetoelastic, and magnetotransport properties in $R$AuGe for $R=$ Dy, Ho, and Gd, and compare them with the nonmagnetic analogue YAuGe.
We observe two-step metamagnetic transitions in DyAuGe and HoAuGe and a metamagnetic transition in GdAuGe for $H\parallel c$.
Their metamagnetism is closely related to the magnetostriction in the $ab$ plane, suggesting an unconventional magnetic structure induced by the magnetic frustration.
Electrical transport measurements reveal magnetoresistance due to the metamagnetism and identify AHE scales with field-induced magnetization.
We also perform the physical property measurements for GdAuGe and observe behavior similar to that reported in the previous study \cite{ram2023multiple}.
In contrast to DyAuGe and HoAuGe, the magnetostriction is monotonic in fields, which is attributed to the absence of the crystal electric field (CEF) effect in Gd$^{3+}$ ions.
We observe enhanced anomalous magnetotransport responses with magnetoresistance reaching 345\% at $T=1.8$ K and $\mu_0 H=9$ T, which is enabled by a higher mobility of hole-type carriers.
These results provide insight into the coupling between frustrated magnetism and high-mobile conduction electrons in the magnetic polar semimetals $R$AuGe.

\section{Experimental Methods}
Single crystals were grown by using Au-Ge self-flux as described in Ref. \cite{kurumaji2023single}.
X-ray crystallographic analyses of single crystals of $R$AuGe were performed using synchrotron radiation of SPring-8 at the BL02B1 beamline.
The x-ray wavelength was 0.31079 \AA.
He-gas-blowing device was employed for controlling the temperature between 30 and 300 K.
The magnetization ($M$) was measured with a superconducting quantum interference device magnetometer (Quantum Design MPMS-XL).
Electrical transport measurements were performed with a conventional five-probe method at a typical frequency near 37 Hz.
The transport properties at low temperatures in a magnetic field were measured with a commercial superconducting magnet and cryostat.
The obtained longitudinal and transverse resistivities were field-symmetrized and antisymmetrized, respectively, to correct contact misalignment.
Thermal expansion and magnetostriction ($\Delta L(T,H)/L_0$) were measured by the fiber Bragg grating (FBG) technique using an optical sensing instrument (Hyperion si155, LUNA) in an Oxford Spectromag.
The change in lattice length $\Delta L(T,H)=L(T,H)-L_0$ from the origin $L_0$ at $T=1.8$ K and $\mu_0 H=0$ T was measured at fixed $H$ and $T$ for thermal expansion and magnetostriction, respectively.
Optical fiber was glued with epoxy (Stycast1266) to the (001) surfaces of as-grown crystals to measure the elongation/compression along the $ab$-plane.

\section{Results}
\subsection{Resistivity in RAuGe and magnetoresistance in YAuGe}
Figures \ref{figreshall}(b) and (c) show the temperature dependence of the zero-field lattice constants ($a$ and $c$ axes) and the resistivity ($\rho_{xx}$) in $R$AuGe ($R=$ Dy, Ho, Gd, and Y).
For all compounds, the lattice shrinks with decreasing the temperature and no signature of symmetry lowering can be seen down to $T=30$ K.
The $\rho_{xx}$ decreases with decreasing temperature, consistent with semimetallic band structure with finite density of states at the Fermi energy \cite{schnelle1997crystal}.
The residual resistivity ratio, defined as the ratio of the resistivities at $T=300$ K to that at $T=1.8$ K ($\rho_{300\text{ K}}/\rho_{1.8\text{ K}}$), are 2.3, 2.6, 18.6, and 4.0 for $R=$ Dy, Ho, Gd, and Y, respectively.
These values are comparable to the previous studies \cite{gibson2001crystal,gibson1996susceptibility,ram2023multiple}.
$R=$ Y shows no signature of transition down to 1.8 K, while $R=$ Dy, Ho, and Gd show an anomaly at low temperatures corresponding to the antiferromagnetic transition temperatures ($T_{\text{N1}}$) as shown in Fig. \ref{figreshall}(d), suggesting the electron-spin couplings.

\begin{figure}[t]
	\includegraphics[width =  \columnwidth]{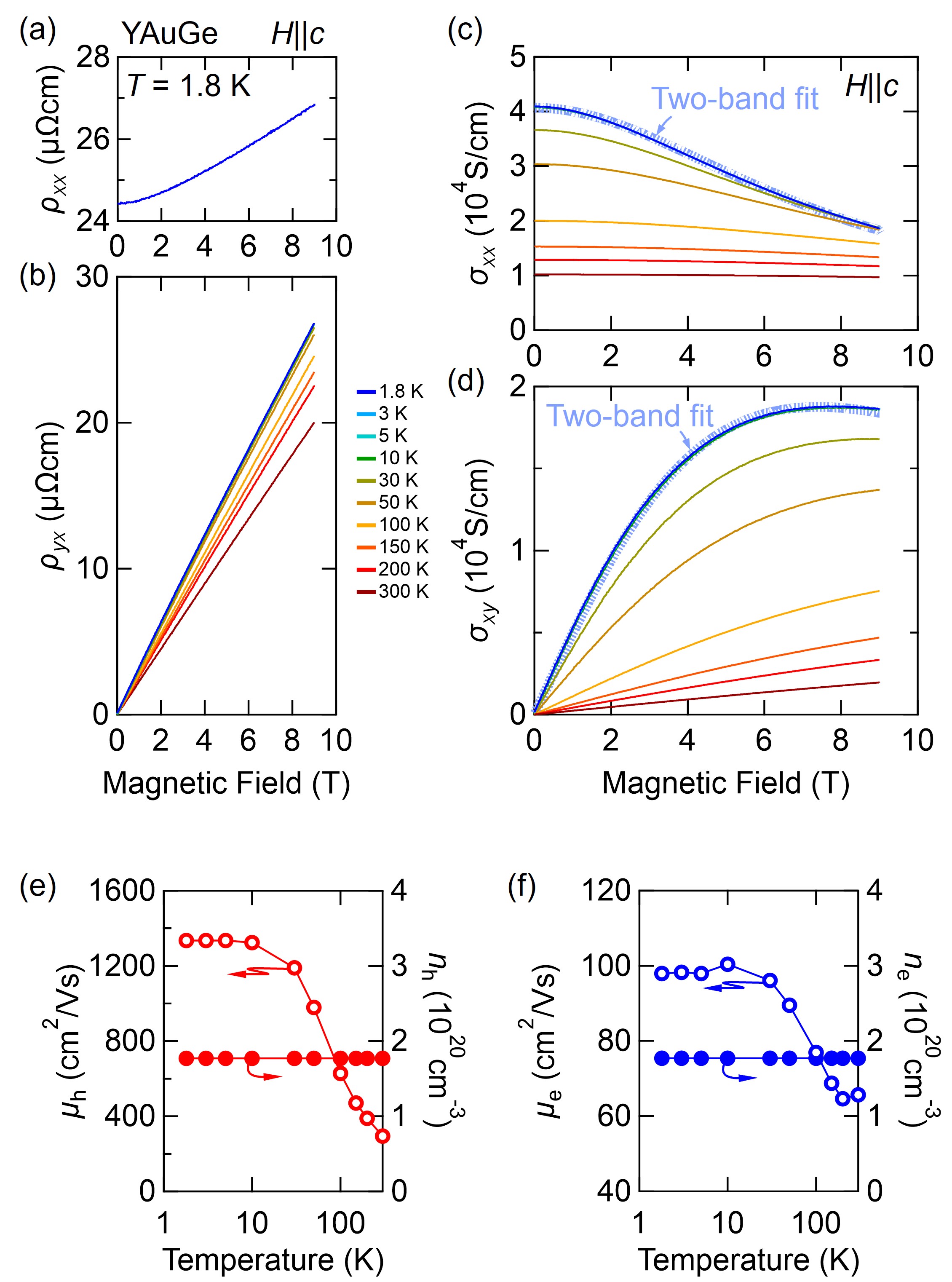}
	\caption{\label{figyag} (a)-(b) Magnetic-field dependence of (a) $\rho_{xx}$ at $T=1.8$ K and (c) the Hall resistivity ($\rho_{yx}$) for $H\parallel c$ at various temperatures in YAuGe.
 (c)-(d) Magnetic-field dependence of (c) $\sigma_{xx}$ and (d) $\sigma_{xy}$ at various temperatures, calculated by Eqs. (\ref{sxxrxx})-(\ref{sxyryx}).
 Dotted pale blue curves are the results of the two-band fit of the data at $T=1.8$ K, using Eqs. (\ref{TBsxx})-(\ref{TBsxy}).
 (e)-(f) Two-band fitting parameters at each temperature.
 The carrier densities are fixed to constant and $n_{\text{e}}=n_{\text{h}}$.
 }
\end{figure}

As a reference of a nonmagnetic compound, we measured the magnetotransport property in YAuGe for $H\parallel c$ (Fig. \ref{figyag}(a)-(b)).
$\rho_{xx}$ shows a typical parabolic positive magnetoresistance consistent with semimetallic electron band structure.
$\rho_{yx}$ shows a positive slope, indicating the hole-type carriers as dominant conduction charges \footnote{In the previous study on GdAuGe \cite{ram2023multiple}, the positive slope of $\rho_{xy}(=-\rho_{yx})$ was reported, which was attributed to the hole-type carrier conduction. In the present study, we adopt the conventional right-handed coordinate system, and a linear response equation: $E_i=\rho_{ij}J_j$ to relate the electric field $E_i$ and the electric current $J_i$ ($i,j=x,y,z$). This leads to $\rho_{yx}>0$ for the hole-type conduction.}.
The slope varies with temperature as the mobility is sensitive to the scattering mechanisms, indicating the multi-carrier transport in YAuGe.
This is consistent with the previous band calculations \cite{pottgen1996crystal,schnelle1997crystal}.

In order to verify the multi-carrier nature, $\rho_{xx}$ and $\rho_{yx}$ are converted to conductivity $\sigma_{xx}$ and Hall conductivity $\sigma_{xy}$ as shown in Figs. \ref{figyag}(c)-(d), by using the following formulas.
\begin{equation}\label{sxxrxx}
    \sigma_{xx}=\frac{\rho_{yy}}{\rho_{xx}\rho_{yy}+\rho_{yx}^2}=\frac{\rho_{xx}}{\rho_{xx}^2+\rho_{yx}^2},
\end{equation}
\begin{equation}\label{sxyryx}
    \sigma_{xy}=\frac{\rho_{yx}}{\rho_{xx}\rho_{yy}+\rho_{yx}^2}=\frac{\rho_{yx}}{\rho_{xx}^2+\rho_{yx}^2}.
\end{equation}
At the last equality for each equation, we assume the isotropic condition $\rho_{yy}=\rho_{xx}$, which is reasonable in the current hexagonal system in the out-of-plane field $H\parallel c$ \cite{kurumaji2023symmetry}. 

We simultaneously fit the $\sigma_{xx}$ and $\sigma_{xy}$ with the formulas for the two-band model below.
\begin{equation}\label{TBsxx}
    \sigma_{xx}=\frac{en_{\text{h}}\mu_{\text{h}}}{1+(\mu_{\text{h}}B)^2}+\frac{en_{\text{e}}\mu_{\text{e}}}{1+(\mu_{\text{e}}B)^2},
\end{equation}
\begin{equation}\label{TBsxy}
    \sigma_{xy}=\frac{en_{\text{h}}\mu^2_{\text{h}}B}{1+(\mu_{\text{h}}B)^2}-\frac{en_{\text{e}}\mu^2_{\text{e}}B}{1+(\mu_{\text{e}}B)^2},
\end{equation}
where $n_{\text{h}}$ and $n_{\text{e}}$ are the hole and electron carrier densities, and $\mu_{\text{h}}$ and $\mu_{\text{e}}$ are the hole and electron mobilities.
The magnetic field $B$ is expressed by a summation as $B=\mu_0 H_{\text{ext}}-N_{\text{d}}M+M$, where $H_{\text{ext}}$ is the external magnetic field, $N_{\text{d}}$ is the demagnetization factor of crystals.
Based on the charge-balanced chemical formula, we adopt the compensated condition $n_{\text{e}}=n_{\text{h}}$ and assume for simplicity that they remain constant as a function of temperature \footnote{If we allow $n_{\text{e}}$ to vary freely, it interferes with $\mu_{\text{e}}$ and becomes two orders of magnitude higher than $n_{\text{h}}$.
This is incompatible with the semimetallic electronic structure in YAuGe.
Even if we allow for the temperature dependence of the carrier densities, the variation is within 10\%.}.

The representative fitting results at $T=1.8$ K are shown by the dotted curves in Figs. \ref{figyag}(c)-(d).
The fitting parameters at various temperatures are summarized in Figs. \ref{figyag}(e)-(f) and capture a systematic suppression of the mobilities with increasing temperature due to electron-phonon scattering.
We note that $\mu_{\text{h}}$ is an order of magnitude higher than $\mu_{\text{e}}$, consistent with the hole-dominant carrier transport.

\subsection{Metamagnetic transition in the c-axis field in RAuGe}
\begin{figure}[t]
	\includegraphics[width =  \columnwidth]{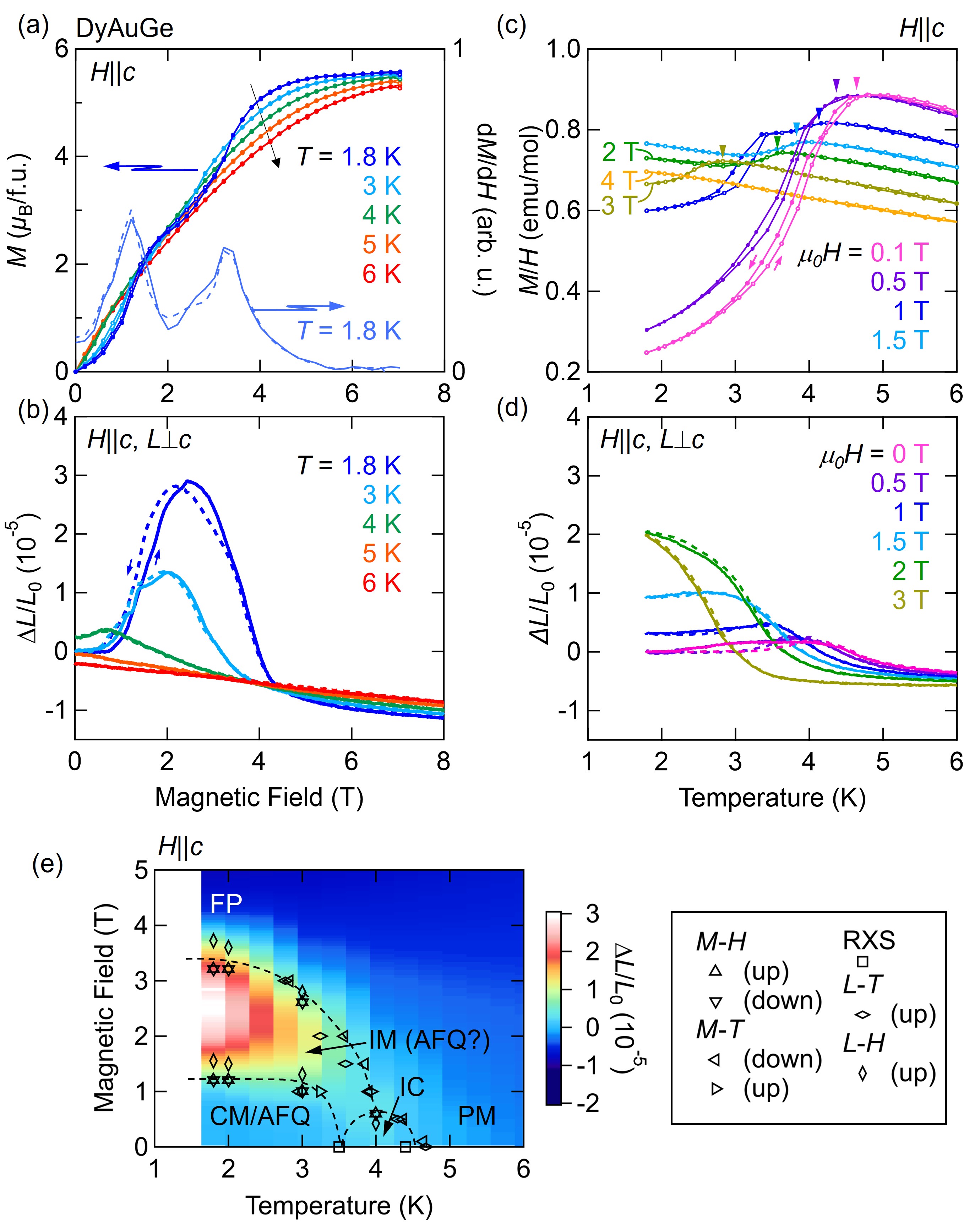}
	\caption{\label{figdyagmh} (a)-(b) Magnetic-field dependence of (a) $M$ and (b) $\Delta L/L_0 =(L(T,H)-L(1.8 \text{ K},0 \text{ T}))/L(1.8 \text{ K},0 \text{ T})$ for $H\parallel c$ at various temperatures in DyAuGe.
 The solid (dashed) curve is for a field increase (decrease) process.
 In (a), the field-derivative of $M$ (d$M$/d$H$) at $T=1.8$ K is also shown.
 (c) Temperature dependence of $M/H$ for $H\parallel c$ with various magnetic fields.
 An open (closed) symbol indicates a temperature increase (decrease) process.
 (d) Temperature dependence of $\Delta L/L_0$ for $H\parallel c$ with various magnetic fields.
 The solid (dashed) curve is for the temperature decrease (increase) process.
 (e) $H-T$ phase diagram for $H\parallel c$ determined by physical property measurements. PM: paramagnetic; IC: incommensurate; CM/AFQ: commensurate antiferromagnetic antiferroquadrupolar; IM: intermediate metamagnetic; FP: field-induced polarized phases, respectively.
 The color map of $\Delta L/L_0$ is also shown.
}
\end{figure}

\begin{figure}[t]
	\includegraphics[width =  \columnwidth]{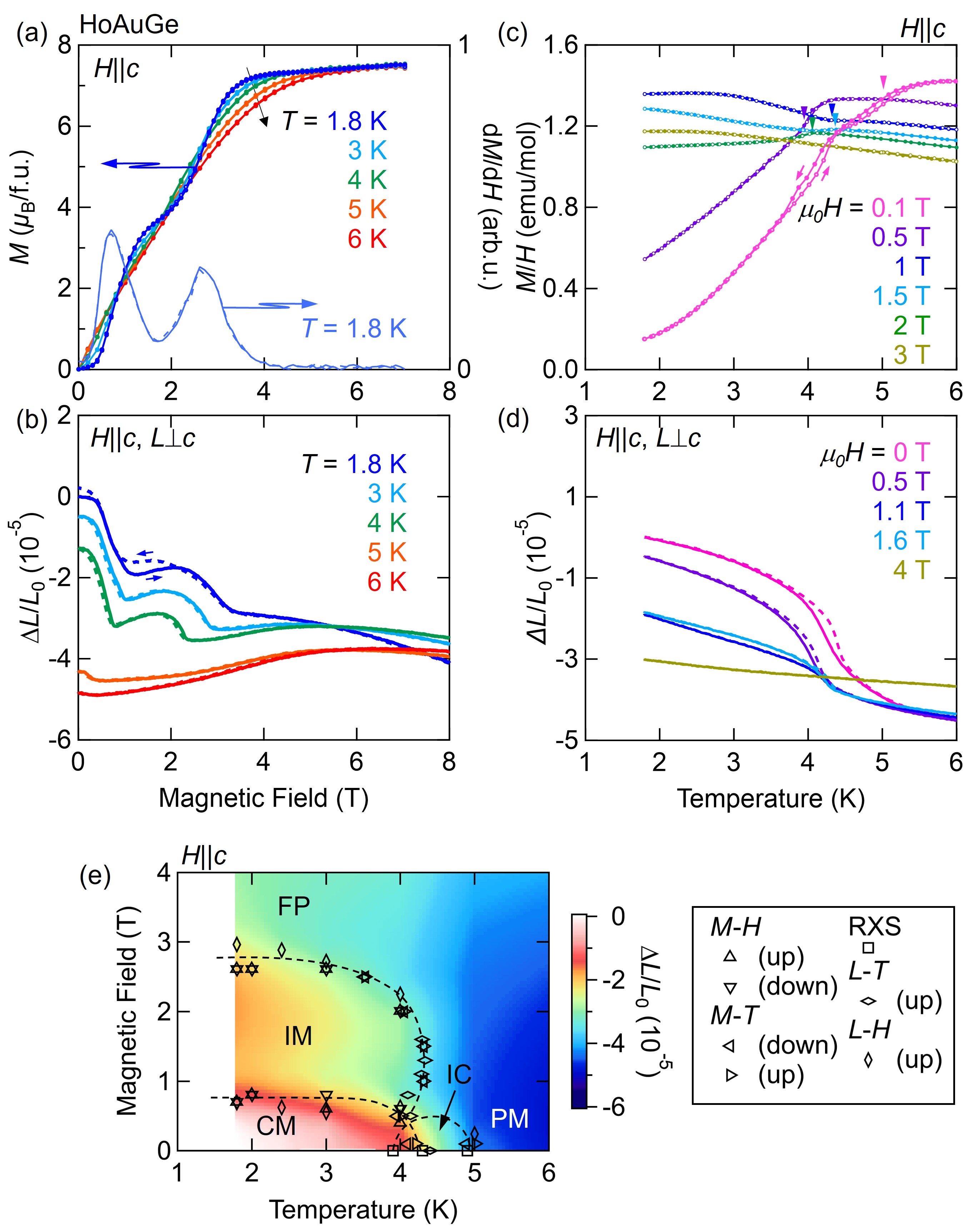}
	\caption{\label{fighoagmh} 
Magnetic and magnetoelastic properties for HoAuGe.
(a)-(e) correspond to Figs. \ref{figdyagmh}(a)-(e) for DyAuGe.
}
\end{figure}

\begin{figure}[t]
	\includegraphics[width =  \columnwidth]{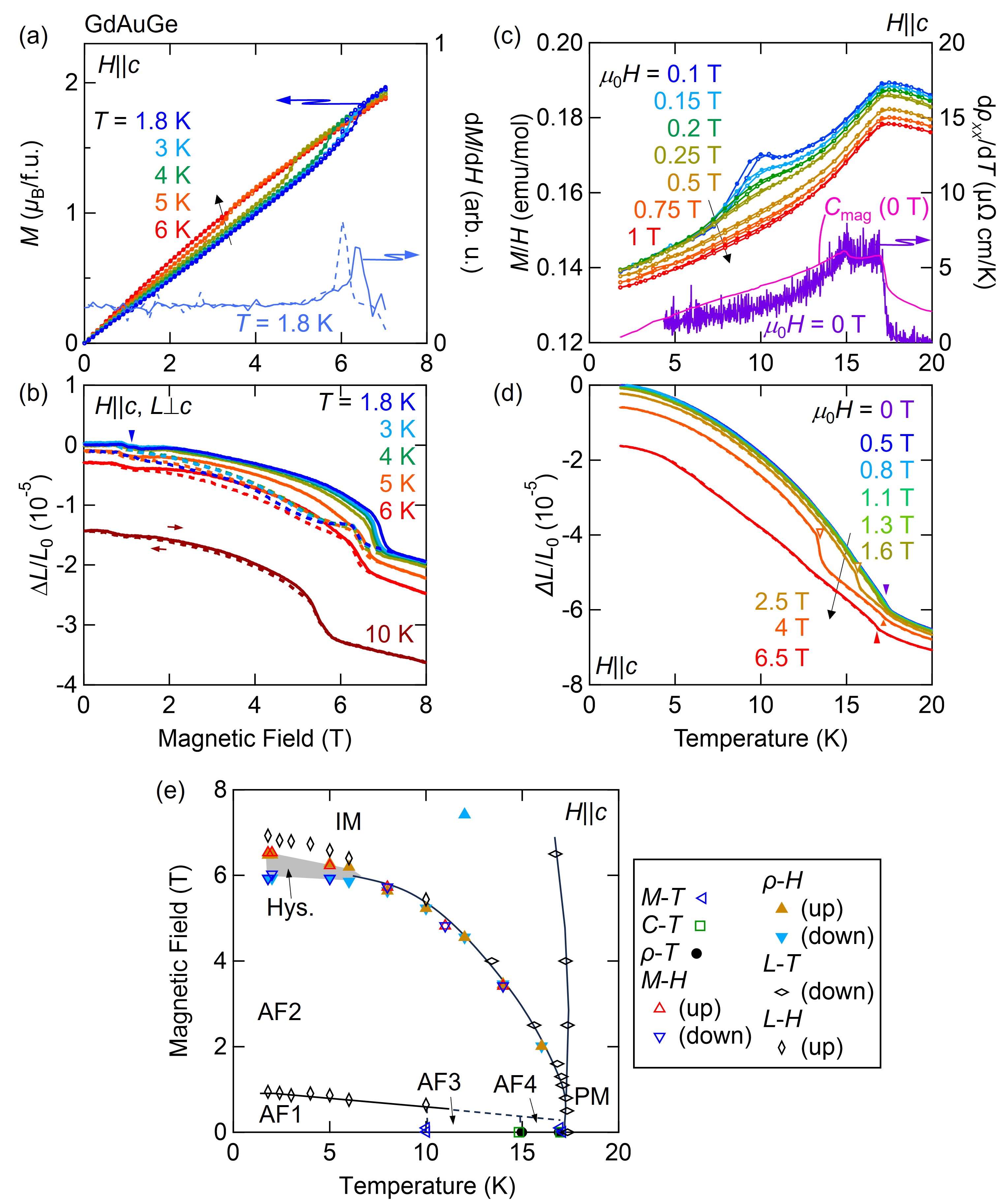}
	\caption{\label{figgdagmh}
 Magnetic and magnetoelastic properties for GdAuGe.
 (a)-(d) correspond to Figs. \ref{figdyagmh}(a)-(d) for DyAuGe.
 In (c), the temperature-derivative of $\rho_{xx}$ ($\text{d}\rho_{xx}/\text{d}T$) at zero field and the magnetic part of the specific heat ($C_{\text{mag}}$) are reproduced from Ref. \onlinecite{kurumaji2023single} are also shown.
 $C_{\text{mag}}$ is scaled to the peaks of $\text{d}\rho_{xx}/\text{d}T$.
 (e) $H-T$ phase diagram for $H\parallel c$ determined by physical property measurements.
 PM: paramagnetic; AF1-AF4: antiferromagnetic; IM: intermediate metamagnetic phases, respectively. 
 The gray hatched area is the hysteresis region observed in the $M$-scan process.
 }
\end{figure}

Before discussing the anomalous magnetotransport properties in $R$AuGe for $R=$ Dy, Ho, and Gd, we first examine the magnetic properties in an out-of-plane magnetic field ($H\parallel c$).
Figures \ref{figdyagmh}, \ref{fighoagmh}, and \ref{figgdagmh} summarize the magnetization and magnetostriction measurements with $H\parallel c$ for DyAuGe, HoAuGe, and GdAuGe, respectively.

Figures \ref{figdyagmh}(a)-(d) are the field and temperature dependence of $M$ and $\Delta L/L_0$.
As shown in Fig. \ref{figdyagmh}(a), $M$ at $T=1.8$ K has a two-step increase with an intermediate metamagnetic (IM) state at around 2 T, suggesting the magnetic frustration in the triangular lattice of Dy.
The field-derivative of $M$ at $T=$ 1.8 K clarifies the two critical fields at about 1.2 T and 3.2 T, respectively.
For a free Dy$^{3+}$ ion with $J=15/2$, the fully polarized magnetic moment is expected to be $M=10$ $\mu_{\text{B}}$/f.u., when the ground state wave function is represented by the quantum number $J_z=\pm 15/2$.
The observed saturation moment at $\mu_0 H=7$ T is 5.6 $\mu_{\text{B}}$/f.u., suppressed from this expectation, which is attributed to the CEF effect giving the ground state wave functions different from the pure $J_z=15/2$.

The metamagnetism can also be seen in the magnetostriction, which is characterized by an elongation of the $a$ or $b^*$ axis in the IM phase (Fig. \ref{figdyagmh}(b)).
We note that the zero-field thermal expansion (Fig. \ref{figdyagmh}(d)) does not show a large evolution of the order of 10$^{-4}$ as reported in Ref. \cite{kurumaji2024canted}, possibly due to the cancellation of the anisotropic lattice deformation by a multi-domain formation under the optical fiber in this measurement setting.
As the out-of-plane field does not lift the degeneracy of the domains, the elongation in the IM state can be attributed to the isotropic expansion of the lattice.
In contrast, the field-induced polarized (FP) state at higher fields (Fig. \ref{figdyagmh}(b)) shows a negative slope.

This nonmonotonic behavior in the magnetostriction can be understood in the context of a competition between exchange striction and the CEF effect \cite{doerr2005magnetostriction}.
The exchange-striction mechanism is attributed to changing $R$-$R$ bond distance to gain exchange interaction energy.
The CEF effect is due to the multipole moment in the 4f electrons of an $R$ ion, which leads to lattice deformation by anisotropic charge distribution.
As seen in the context of GdAuGe (see Fig. \ref{figgdagmh}), the exchange-striction mechanism between Gd ions gives a negative magnetostriction as the magnetic moments evolve toward the applied magnetic field.
Given the nearly identical crystal and electronic structures, the exchange interactions as a function of lattice spacing in $R$AuGe are anticipated to remain independent of $R$, explaining the observed negative slope of $\Delta L/L_0$ (Fig. \ref{figdyagmh}(b)) in the FP state for DyAuGe.
The positive magnetostriction in the IM state, on the other hand, cannot be attributed to the exchange striction mechanism, and may be due to the CEF effect.

In the previous study \cite{kurumaji2024canted}, it was shown that the orientation of the magnetic dipole moments in Dy$^{3+}$ ions is strongly coupled to the electric quadrupolar moments, which is due to a quasi-quartet nature of the CEF levels.
The canted magnetic structure in the IM state is expected to induce a ferroquadrupole ordering, which deforms the lattice in the opposite direction to the exchange striction mechanism.
We also note that the nonmonotonic magnetostriction differs from the behavior in ErGa$_2$, which has a similar IM state in $H\parallel c$ with the three-sublattice collinear state \cite{markin1998magnetoelastic}.
This may indicated that the IM state in DyAuGe is different from a simple collinear state, but has a noncollinear/noncoplanar nature.
Further diffraction studies are a subject of future studies to clarify the magnetic structure in the IM state.

In the previous resonant x-ray diffraction experiment \cite{kurumaji2024canted}, we have revealed the successive magnetic transitions in DyAuGe at zero field.
As the temperature decreases from the paramagnetic state, DyAuGe shows a transition at $T_{\text{N1}}=4.4$ K to the incommensurate magnetic modulation state with the modulation vector $q=(0.398, 0, 0)$ in the reciprocal lattice unit.
The commensurate modulation with $q=(1/2, 0, 0)$ starts to develop below $T_{\text{N2}}=3.5$ K, which is identified as a canted antiferromagnetic state associated with an antiferroquadrupole order \cite{kurumaji2024canted}.
Figure \ref{figdyagmh}(c) shows the evolution of the antiferromagnetic transition temperature under $H\parallel c$.
The transition from the paramagnetic state to a magnetically ordered state can be identified as marked by triangles, which shift to a lower temperature with increasing field.
The transition anomalies are summarized in the $H-T$ phase diagram (Fig. \ref{figdyagmh}(e)), where the phase boundary for the IM state is identified.
The color map of the magnetostriction shows that the enhancement of the magnetostriction is clearly correlated with the IM state.
We note that the FP and PM states are continuously connected in the phase diagram, suggesting that the FP is the $q=0$ state for $M$ at each Dy site polarized along the $c$ axis.

We measure the magnetization in HoAuGe for $H\parallel c$ at various temperatures as shown in Fig. \ref{fighoagmh}(a).
$M$ shows a two-step metamagnetic transition below $T< T_{\text{N1}}$.
The saturation value of $M$ at $\mu_0 H=7$ T is 7.5 $\mu_{\text{B}}$/f.u. at $T=1.8$ K, which is suppressed by the CEF effect from the expectation 10 $\mu_{\text{B}}$ for a free Ho$^{3+}$ ion as in the case of DyAuGe.
The transition fields are clearly identified by taking the field derivative of $M$.
The metamagnetism can also be seen in the magnetostriction in the $ab$ plane (Fig. \ref{fighoagmh}(b)).
The $\Delta L/L_0$ shows a dip at a field corresponding to the metamagnetic transitions, suggesting a quadrupolar correlation due to the CEF in the IM state as observed in DyAuGe.
The magnitude of the lattice length change is several times smaller than in DyAuGe, consistent with the less significant quadrupolar moment in HoAuGe \cite{kurumaji2024canted}.

To draw the magnetic phase diagram for $H\parallel c$, we measured the temperature dependence of $M/H$ (Fig. \ref{fighoagmh}(c)) and the thermal expansion $\Delta L/L_0$ (Fig. \ref{fighoagmh}(d)) at various magnetic fields for $H\parallel c$.
At $\mu _0 H = 0.1$ T, a high-temperature anomaly as denoted by the solid triangle in Fig. \ref{fighoagmh}(c) is related with the transition from the paramagnetic state to the incommensurate modulation state, and the low-temperature drop with hysteresis is related to the transition to the commensurate antiferromagnetic state \cite{kurumaji2024canted}.
As the field increases, the high-temperature anomaly moves to a lower temperature and disappears above $\mu_0H = 3$ T, suggesting the entering into the field-induced polarized (FP) state.
Similar trends for the transition temperatures are also seen in $\Delta L/L_0$ (Fig. \ref{fighoagmh}(d)).
We plot the transition anomalies observed in the above measurements and show the $H-T$ phase diagram in Fig. \ref{fighoagmh}(e).
Magnetostriction is also correlated with the magnetic transitions.
The absence of a boundary separating the FP and PM states indicates that the FP state is $q=0$ with $M$ polarized along the $c$ axis.
We note that a similar metamagnetism has been reported in an isostructural (LiGaGe-type) HoAuSn \cite{ueda2023colossal}, which was identified as a by-product in the study of the Heusler-type polymorph.
Exploration of the metamagnetism in LiGaGe-type compounds and clarifying the effect of magnetic frustration would be an interesting subject for future study.

We study the magnetic property in GdAuGe.
The field dependence of $M$ for $H\parallel c$ exhibits a spin-flop-like transition as shown in Fig. \ref{figgdagmh}(a), which is in agreement with the report in Ref. \cite{ram2023multiple}.
The d$M$/d$H$ exhibits a hysteresis, suggesting the first order nature of this transition.
The field-induced transition also induces an anomaly in the lattice length, where the drop of $\Delta L/L_0$ is observed as shown in Fig. \ref{figgdagmh}(b).
The lattice change is monotonically negative in fields in contrast to the behavior in DyAuGe (Fig. \ref{figdyagmh}(b)) and HoAuGe (Fig. \ref{fighoagmh}(b)).
The exchange striction is expected to be the main origin of the magnetostriction in GdAuGe due to the absence of the orbital contribution in the Gd 4f electrons, which explains the CEF effect in DyAuGe and HoAuGe to give rise to the positive lattice change in the IM state.
We also note that the $\Delta L/L_0$ has a small anomaly at about 1 T, as denoted by the solid triangle, which corresponds to the anomaly identified in Ref. \cite{ram2023multiple}.

\begin{figure*}[t]
	\includegraphics[width =  0.8\textwidth]{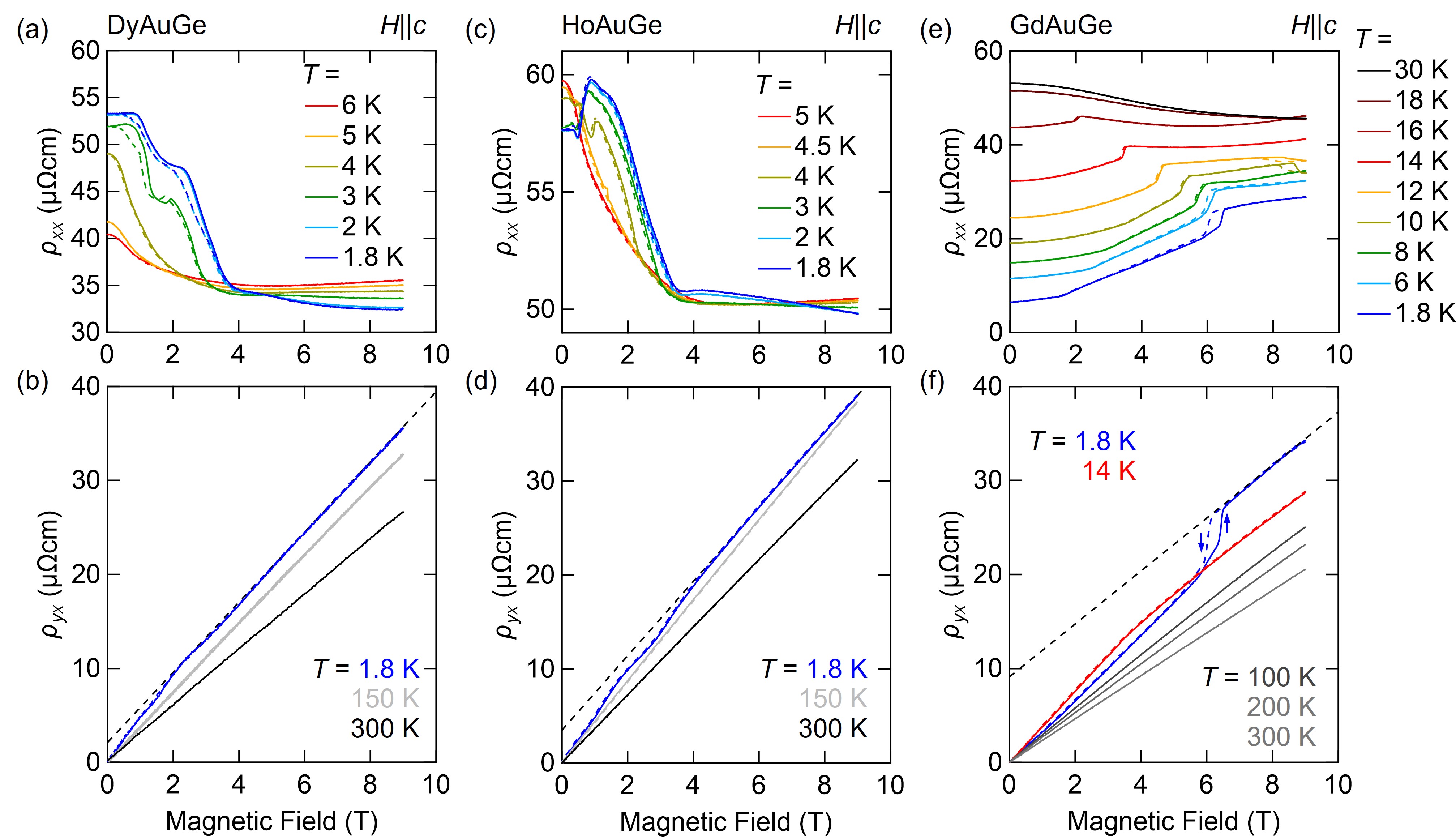}
	\caption{\label{figragrh} (a)-(b) Field dependence of (a) $\rho_{xx}$ and (b) $\rho_{yx}$ for $H\parallel c$ at various temperatures in DyAuGe.
 The solid (dashed) curve is for a field increase (decrease) process.
 In (b), the black dashed line denotes the $B$-linear fit of $\rho_{yx}$ at $T=1.8$ K in the field range $\mu_0 H=7-9$ T.
 The results in (c)-(d) and (e)-(f) correspond to HoAuGe and GdAuGe, respectively.
 }
\end{figure*}

Figure \ref{figgdagmh}(c) shows the temperature dependence of $M/H$.
The transition at $T_{\text{N}}=17$ K for $\mu_0H=0.1$ T is indicated by a peak.
As pointed out in the previous study \cite{kurumaji2023single}, we identified another transition at about $T=10$ K.
As the field increases, this anomaly becomes weaker and almost negligible at $\mu_0 H=1$ T.
Figure \ref{figgdagmh}(c) also shows the temperature derivative of $\rho_{xx}$, which has peaks at about $T=17$ K and 14.8 K.

Figure \ref{figgdagmh}(d) shows the temperature dependence of $\Delta L/L_0$ at various temperatures.
The transition from the PM to a magnetically ordered state can be seen as a kink marked by solid triangles.
For $\mu_0H=2.5$ T and 4 T, another low-temperature kink (open triangle) appears, suggesting an antiferromagnetic-antiferromagnetic transition.
We summarize the anomalies identified in the magnetization and magnetostriction measurements in Fig. \ref{figgdagmh}(e).
Four antiferromagnetic states (AF1-AF4) are identified below the critical field for the intermediate metamagnetic (IM) state.
The separation of two antiferromagnetic states, AF3 and AF4, was not seen in \cite{ram2023multiple}, possibly due to some difference of the sample quality grown in different fluxes \cite{huang2012low}.
We note that the lattice parameters reported in Ref. \cite{ram2023multiple} ($a=b=4.4281$ \AA, $c=7.4262$ \AA) are slightly longer than in our report \cite{kurumaji2023single} ($a=b=4.42799(1)$ \AA, $c=7.4176(2)$ \AA).
Identification of the magnetic structure of these states and possible growth condition dependence are a subject for future study.
We also note that $M$ at the highest magnetic field (Fig. \ref{figgdagmh}(a)) is far below the expected saturation value of 7 $\mu_{\text{B}}$ for a Gd$^{3+}$ ion.
Exploring the phase diagram at higher magnetic fields is also an interesting future study, as the field-induced topological magnetic texture such as magnetic skyrmions are expected to appear in the Gd-based frustrated triangular-lattice magnet \cite{kurumaji2019skyrmion,kurumaji2020spiral}.

\subsection{Anomalous transport properties in RAuGe}
In the previous sections, we have established the magnetic properties of $R$AuGe ($R=$ Dy, Ho, and Gd) and revealed the magnetic phase diagrams for $H\parallel c$.
In this section, we focus on the magnetotransport properties and the AHE.
We summarize in Fig. \ref{figragrh} the field dependence of the transport properties in $R$AuGe ($R=$ Dy, Ho, and Gd).

For DyAuGe, $\rho_{xx}$ (Fig. \ref{figragrh}(a)) shows a step-wise suppression with increasing field, which is related to the metamagnetic transitions (Fig. \ref{figdyagmh}(a)).
The Hall resistivity $\rho_{yx}$ (Fig. \ref{figragrh}(b)) shows a positive slope in $H$ for the hole-type carrier conduction, and the slope becomes smaller at higher temperatures, as observed in YAuGe (Fig. \ref{figyag}(b)). 
At low fields below the magnetization saturation, a small variation of the slope at $T=1.8$ K is observed as a signature of AHE.

Anomalous Hall resistivity in metals is often analyzed using the following ansatz \cite{pugh1953hall}:
\begin{equation}\label{ahe}
    \rho_{yx}=R_0B+R_{\text{s}}M,
\end{equation}
where $R_0$ is the ordinary Hall effect and $R_{\text{s}}$ is the coefficient for the AHE proportional to the $M$.
We fit the high-field slope of $\rho_{yx}$ at $T=1.8$ K with a $B$-linear function, assuming that $R_0$ and $R_{\text{s}}$ are constants.
The intercept at the zero field is a finite positive value ($\rho_{yx0}^{\text{A}}=2.1$ $\mu \Omega$ cm), consistent with the $M$ induced AHE.

We observe a similar behavior in HoAuGe.
As shown in Fig. \ref{figragrh}(c), $\rho_{xx}$ has a clear anomaly across the metamagnetic transitions.
The increase of $\rho_{xx}$ in the IM state is in contrast to that for DyAuGe, which could be related to the different magnetic structures in these compounds.
The Hall resistivity at $T=1.8$ K has step-wise changes in slope at metamagnetic transitions, suggesting the AHE contribution (Fig. \ref{figragrh}(d)).
The linear fit has a finite positive intercept at zero field, corresponding to $\rho_{yx0}^{\text{A}}=3.5$ $\mu \Omega$ cm.

We also investigate magnetotransport properties in GdAuGe.
At $T=1.8$ K, $\rho_{xx}$ increases parabolically in $H$ at low fields and shows a step-wise increase at the spin-flop-like transition (Fig. \ref{figragrh}(e)), which is sharper than that reported in Ref. \cite{ram2023multiple}.
The magnetoresistance becomes negative above $T=18$ K ($>T_{\text{N}}$), indicating the suppression of spin-disorder scattering in the paramagnetic state.
The increase in Hall resistivity reported in Ref. \cite{ram2023multiple} is also reproduced with a larger step at $T=1.8$ K as shown in Fig. \ref{figragrh}(f).
The linear fit at high fields gives an estimate of $\rho_{yx0}^{\text{A}}=6.5$ $\mu \Omega$ cm.
We warn that there is a possibility of the overestimation because the multi-carrier nature leads to the deviation of the $\rho_{yx}-B$ curve from the $B$-linear trend at high fields, where the Hall angle ($\rho_{yx}/\rho_{xx}$) becomes comparable to one.

In the above discussion, we identify a signature of the anomalous Hall response in $R$AuGe ($R=$ Dy, Ho, and Gd).
We note, however, that the linear fit of $\rho_{yx}$ gives only a rough estimate of the anomalous Hall component.
In the case of compensated semimetals, multi-carrier nature potentially deforms the $\rho_{yx}$ from $B$-linear behavior because $R_0$ is a function of mobilities and carrier densities and is no longer expected to be a constant \cite{kurumaji2024metamagnetic}.
In particular, magnetic subsystems affect the relaxation time of conduction carriers, which is not considered in the ansatz (Eq. (\ref{ahe})).
We also note that the prescription given in Ref. \cite{lee2007hidden}, where $R_{\text{s}}=S_{H}\rho_{xx}^2$ with $S_{H}$ as a constant, is inappropriate for the current system because the assumption $\rho_{xx}>>\rho_{yx}$ is no longer valid.

\begin{figure}[t]
	\includegraphics[width =  \columnwidth]{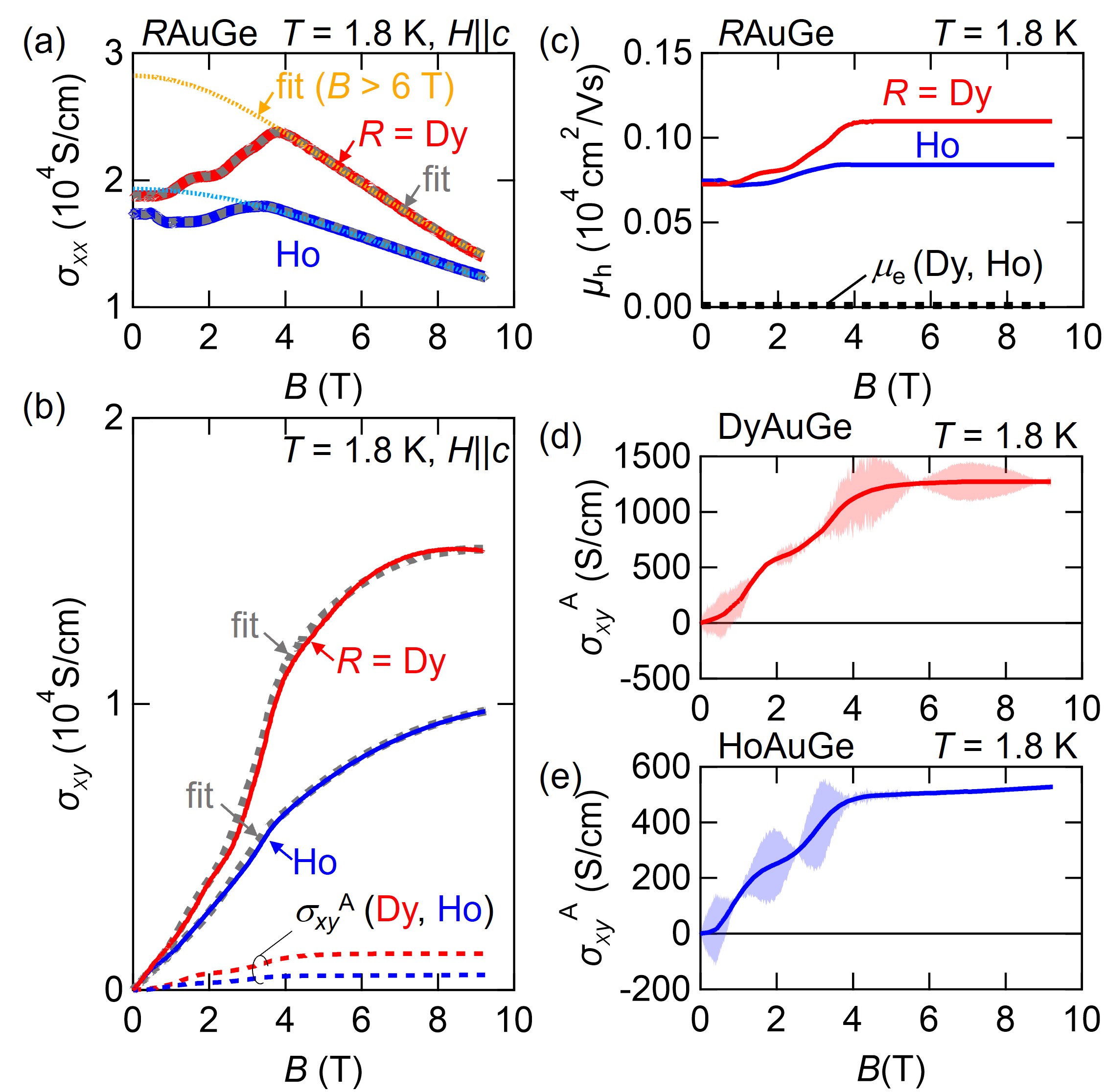}
	\caption{\label{figahe} (a) $B$-dependence of $\sigma_{xx}$ for $R=$ Dy, and Ho.
Solid red (blue) curve is the observed $\sigma_{xx}$ for $R=$ Dy (Ho) converted from $\rho_{xx}$ and $\rho_{yx}$ by using the formula Eq. (\ref{sxxrxx}).
Dotted orange (cyan) curve is the two-band fitting curve (Eq. (\ref{TBsxx})) using the data for $R=$ Dy (Ho) in the high-field region: $B>6$ T and setting the two-band parameters ($n_{\text{e/h}}$, $\mu_{\text{e/h}}$) are constants.
The dashed gray curve is the fit obtained by using the modified two-band model, where the field-dependent $\mu_{\text{h}}$ is introduced in Eq. (\ref{TBsxx}) (see text).
(b) Corresponding $\sigma_{xy}$ for $R=$ Dy (red) and Ho (blue).
The fitting (dashed gray) curves are obtained by using Eq. (\ref{TBsxyA}), which introduces $M$-scaled $\sigma^{\text{A}}_{xy}$ (dashed red (Dy) and blue (Ho) curves).
(c) Field dependence of the hole mobility $\mu_{\text{h}}$ for $R=$ Dy and Ho.
A field-constant electron mobility is also shown by the dashed line.
(d)-(e) Field dependence of $\sigma_{xy}^{\text{A}}$ for (d) $R=$ Dy, and (e) $R=$ Ho.
Error bars correspond to the difference between the fitted curve and the observation in (b).
}
\end{figure}

To analyze the AHE more quantitatively, we discuss the anomalous transport property by $\sigma_{xx}$ and $\sigma_{xy}$.
Figures \ref{figahe}(a)-(b) show the field-dependence of $\sigma_{xx}$ and $\sigma_{xy}$ for $R=$ Dy, and Ho, calculated by using Eqs. (\ref{sxxrxx})-(\ref{sxyryx}).
The field-dependence of $\sigma_{xx}$ (Fig. \ref{figahe}(a)) clearly shows nonmonotonic behavior, which is correlated with the evolution of the magnetization before saturation.
The $\sigma_{xy}$ also deviates from a trend as observed in YAuGe (Figs. \ref{figyag}(b)-(c)).

Using the data at $B>6$ T, we fit $\sigma_{xx}$ with Eq. (\ref{TBsxx}), as shown by the dotted orange and cyan curves for $R=$ Dy and Ho, respectively, in Fig. \ref{figahe}(a).
For both $R=$ Dy and Ho, the observed $\sigma_{xx}$ at $B<6$ T has a large discrepancy from the conventional two-band fit due to the magnetoresistance associated with the magnetization process.
The features in $\sigma_{xy}$ at the corresponding region are also related to this effect.

To accurately extract the AHE contribution, a modification of the two-band model incorporating the above effect is necessary.
It is known that the field dependence of the conductivity tensors in metamagnetic materials can be well reproduced by introducing the field-dependence of the mobility \cite{ye2017electronic,kurumaji2024metamagnetic}.
As the hole-type carriers dominate the electrical transport in the current system, we introduce the field-dependence of $\mu_{\text{h}}$ to reproduce that of $\sigma_{xx}$.
The obtained field-dependence of the $\mu_{\text{h}}$ in DyAuGe and HoAuGe is shown in Fig. \ref{figahe}(c).
We note that the electron-type carrier has negligibly small mobility compared to $\mu_{\text{h}}$.
Using these mobilities, we show the simulated $\sigma_{xx}$ for DyAuGe and HoAuGe in Fig. \ref{figahe}(a) by dashed gray curves, which agree well with the observation.

\begin{figure}[t]
	\includegraphics[width =  \columnwidth]{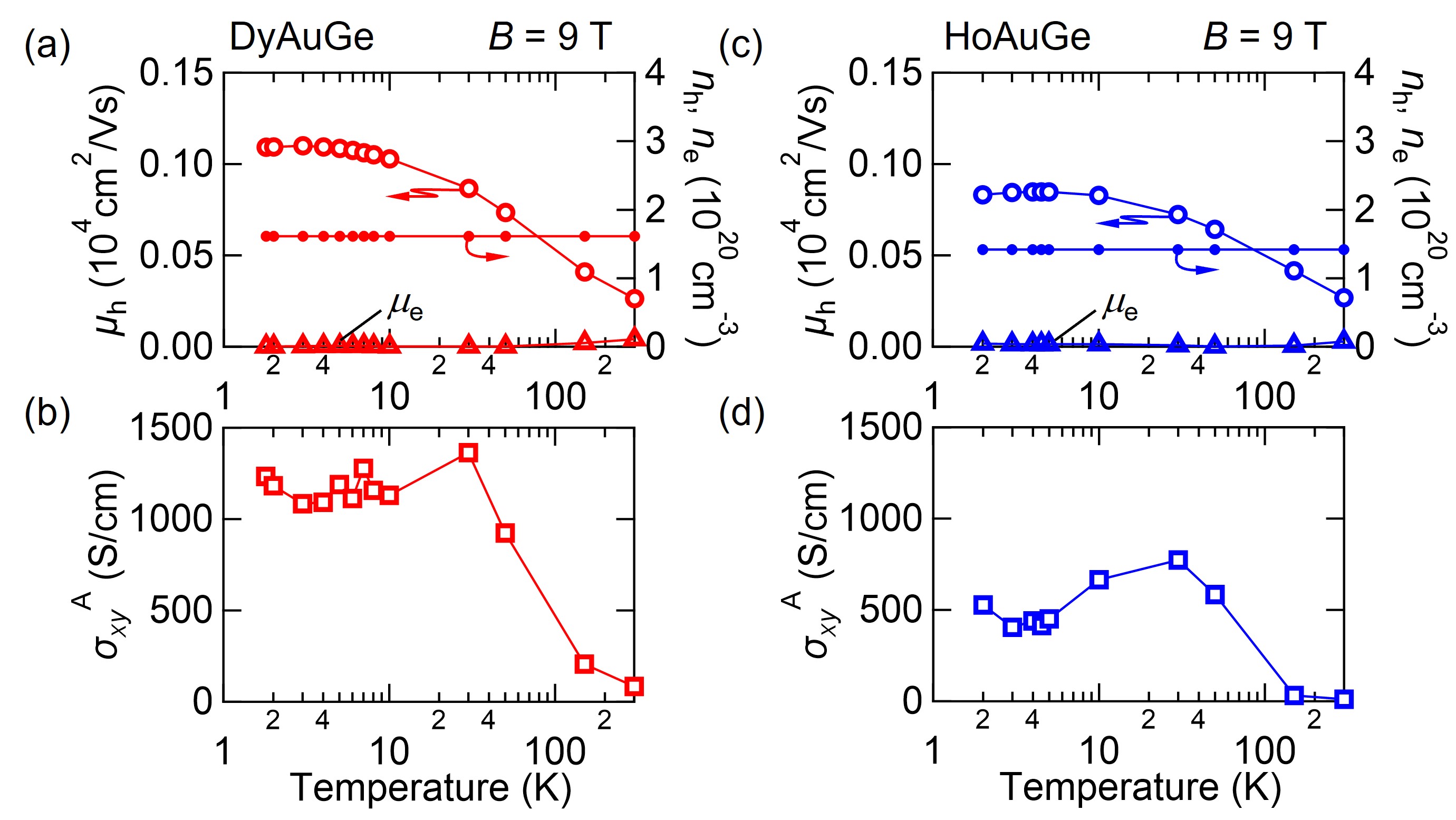}
	\caption{\label{figaheparap} (a)-(b) Modified two-band model parameters and $\sigma_{xy}^{\text{A}}$ at $B=9$ T at each temperature for DyAuGe.
 $n_{\text{e}}$ is set to be identical to $n_{\text{h}}$.
 (c)-(d) Corresponding plots for HoAuGe.
}
\end{figure}

Within the above assumptions, we simultaneously fit the $\sigma_{xx}$ and $\sigma_{xy}$ for DyAuGe and HoAuGe by the two-band model Eqs. (\ref{TBsxx})-(\ref{TBsxy}) with $\mu_{\text{h}}$ shown in Fig. \ref{figahe}(c).
We use the following formula for $\sigma_{xy}$ instead of Eq. (\ref{TBsxy}).
\begin{equation}\label{TBsxyA}
    \sigma_{xy}=\frac{en_{\text{h}}\mu^2_{\text{h}}B}{1+(\mu_{\text{h}}B)^2}-\frac{en_{\text{e}}\mu^2_{\text{e}}B}{1+(\mu_{\text{e}}B)^2}+S_{\text{H}}M.
\end{equation}
The last term is the anomalous Hall conductivity $\sigma_{xy}^{\text{A}}$ scaled to the $M$.
We adopt the rigid band approximation, where the band structure near the Fermi energy is not significantly modified from that of YAuGe.
This is reasonable since the 4f electrons are expected to be well localized at Dy and Ho ions.
The condition $n_{\text{e}}=n_{\text{h}}$ for the compensated semimetal is also introduced to reduce the number of free parameters.

The fitting results for $\sigma_{xy}$ at $T=1.8$ K are shown in Fig. \ref{figahe}(b), where they satisfactorily reproduce the nonmonotonic field-dependence in DyAuGe and HoAuGe.
The $\sigma_{xy}^{\text{A}}$ is shown in Figs. \ref{figahe}(d)-(e), where $\sigma^{\text{A}}_{xy}\sim 1200$ S/cm and $\sim 530$ S/cm at 9 T for DyAuGe and HoAuGe, respectively.
The magnitude of $\sigma_{xy}^{\text{A}}$ at the field-induced magnetization saturation is consistent with the rough estimate from the zero-field intercept of the $\rho_{yx}-H$ curve (Figs. \ref{figragrh}(b) and (d)), i.e., $\sigma_{xy0}^{\text{A}}\sim \rho^{\text{A}}_{yx0}/(\rho^2_{xx}+\rho^2_{yx})$.

We note that the uncertainty of the fit can be estimated by the deviation between the observed $\sigma_{xy}$ and the fitting curves, as shown by the error bars in Figs. \ref{figahe}(d)-(e).
The errors are within $< 400$ S/cm for DyAuGe and $< 150$ S/cm for HoAuGe, which are small compared to the magnitude of the $\sigma_{xy}^{\text{A}}$ at $B=9$ T.
Without the contribution of the $\sigma_{xy}^{\text{A}}$, i.e., setting $S_{\text{H}}=0$, the discrepancy of the fits is enhanced up to $680$ S/cm for DyAuGe and $250$ S/cm for HoAuGe, verifying the relevance of the anomalous Hall component.
We note that the additional component in the AHE such as a topological Hall effect \cite{taguchi2001spin} is not considered in the current analysis.
It is possible if a noncoplanar spin structure is realized in the IM state, while the contribution to $\sigma_{xy}^{\text{A}}$ is expected to be comparable to the error.

Figures \ref{figaheparap}(a)-(b) summarize the fitting parameters at each temperature for $R=$ Dy and Ho.
We note that $n_{\text{h}}$ (=$n_{\text{e}}$) is fixed as a temperature-independent constant.
For both $R=$ Dy and Ho, $\mu_{\text{e}}$ is negligibly small ($<40$ cm$^2$/Vs for $R=$ Dy, $<30$ cm$^2$/Vs for $R=$ Ho), compared to $\mu_{\text{h}}$, and $\mu_{\text{h}}$ is suppressed with increasing temperature.
These features are consistent with the behavior of those in YAuGe (Fig. \ref{figyag}(e)).
$\sigma_{xy}^{\text{A}}$ at low temperatures is comparable to a typical AHE due to the intrinsic origin \cite{onoda2008quantum}.
This suggests that the Berry curvature in the electron bands contributes to the microscopic mechanism.

Noncentrosymmetric structure is expected to host nontrivial band crossing induced by spin-orbit coupling \cite{xu2017discovery,bzduvsek2016nodal,xie2021kramers,zhang2023kramers}.
We note that the space group $P6_3mc$ of the current compounds belongs to the polar and achiral system, where Kramers nodal lines along $k_z$ direction in the momentum space are expected \cite{xie2021kramers}.
The magnetic moment along the $c$ axis lifts the degeneracy by breaking the time-reversal symmetry to give rise to a finite Berry curvature maximized when the Fermi energy is close to the gap \cite{culcer2003anomalous,dugaev2005anomalous,onoda2006intrinsic}.
This might contribute to the AHE in $R$AuGe.
We also note that the extrinsic mechanism has not been excluded \cite{kondo1962anomalous} for the system with rare-earth ions, which is characterized by strong spin-orbit coupling and anisotropic electronic distribution.
Further study of the electronic structure of $R$AuGe is desirable to reveal the origin of the AHE.

\begin{figure}[t]
	\includegraphics[width =  \columnwidth]{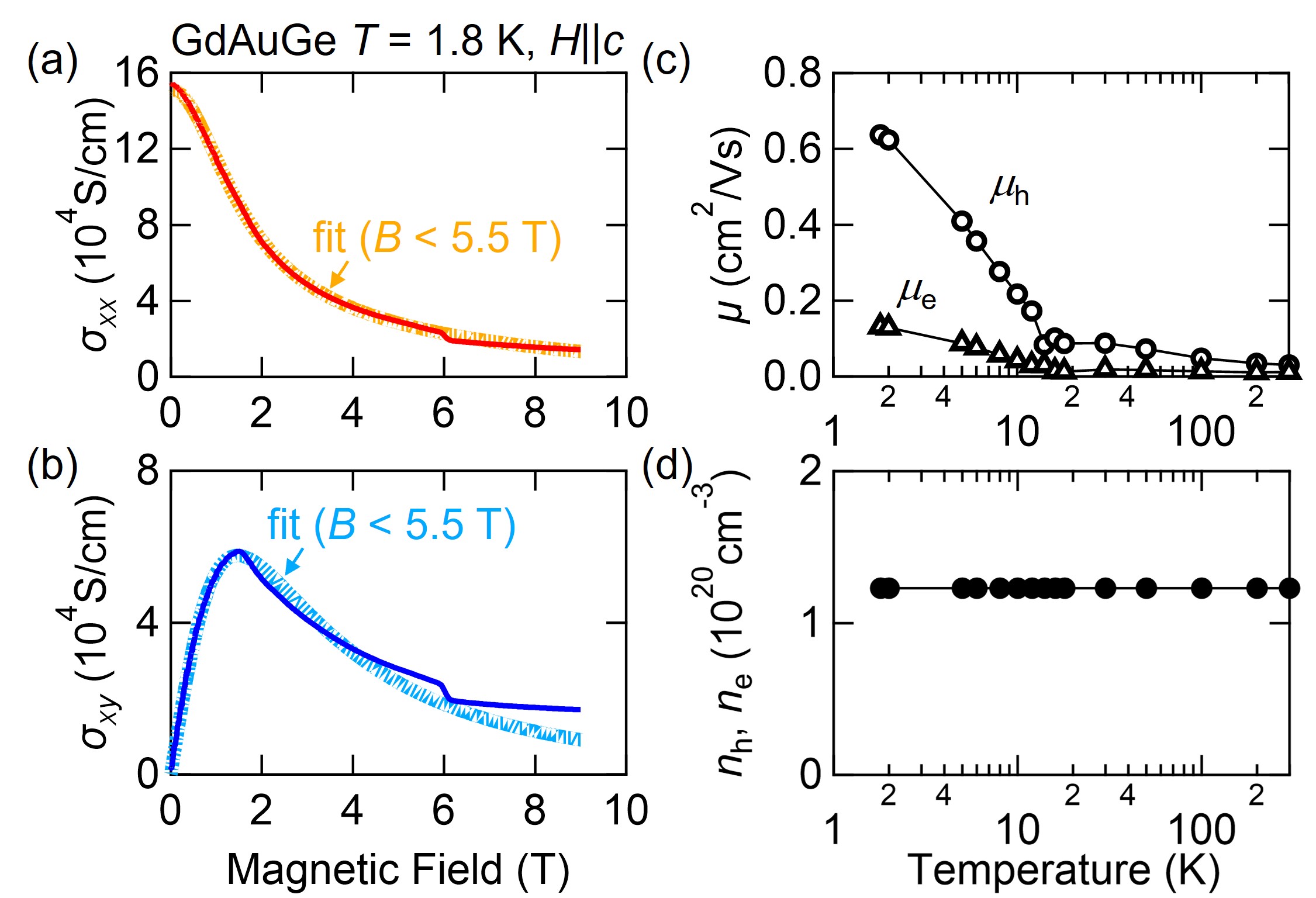}
	\caption{\label{figgdagahe}(a)-(b) Magnetic-field dependence of (a) $\sigma_{xx}$ and (b) $\sigma_{xy}$ for $H\parallel c$ at $T=1.8$ K in GdAuGe.
 Dotted orange and cyan curves are the results of the two-band fitting with Eqs. (\ref{TBsxx})-(\ref{TBsxy}) using the data in $B<5.5$ T. 
 (c)-(d) Two-band fitting parameters at each temperature: (c) $\mu_{\text{h}}$ and $\mu_{\text{e}}$, (d) $n_{\text{h}}$ ($=n_{\text{e}}$), which is fixed at constant.
 }
\end{figure}

Finally, we briefly discuss about the AHE in GdAuGe.
We convert the resistivity tensors into the conductivity tensors as shown in Figs. \ref{figgdagahe}(a)-(b).
The demagnetization effect is not considered because the field-induced magneization is small up to 9 T.
The overall feature in both $\sigma_{xx}$ and $\sigma_{xy}$ is consistent with the multi-carrier behavior with high-mobile hole-type carriers.

The spin-flop-like transition gives a drop in both $\sigma_{xx}$ and $\sigma_{xy}$ at about 6 T.
In the rough estimate using the Hall resistivity (Fig. \ref{figragrh}(f)), the $\sigma_{xy0}^{\text{A}}$ is positive, which is the opposite trend to the step-wise anomaly observed in the $\sigma_{xy}$ curve (Fig. \ref{figgdagahe}(b)).
This means that the possible increase of $\sigma_{xy}$ associated with the AHE effect at the spin-flop transition is masked by the changes in mobility or carrier density.
In order to quantitatively extract the AHE in GdAuGe, it is necessary to clarify the magnetic structures and the associated electronic structure changes.

In order to capture the multi-carrier nature in GdAuGe, we simultaneously fit $\sigma_{xx}$ and $\sigma_{xy}$ with Eqs. (\ref{TBsxx})-(\ref{TBsxy}) by using the data below $B<5.5$ T.
All the parameters are set to be field-independent.
$n_{\text{h}}$ and $n_{\text{e}}$ are fixed to be identical and a temperature-independent constant.
The representative fitting result for $T=1.8$ K is shown by the dotted curves in Figs. \ref{figgdagahe}(a)-(b).
The discrepancy with the observation is enlarged in $\sigma_{xy}$ with increasing field, suggesting a presence of additional factors affecting the magnetotransport in GdAuGe.
The fitting parameters at each temperature are shown in Figs. \ref{figgdagahe}(c)-(d).
$\mu_{\text{h}}$ and $\mu_{\text{h}}$ decay rapidly across $T_{\text{N}}$ due to the effect of spin-electron scattering. 
We note that the $\mu_{\text{e}}$ is smaller than $\mu_{\text{h}}$, but of a comparable order of magnitude, in contrast to the negligible mobility of the electron-type carriers in DyAuGe and HoAuGe (Figs. \ref{figaheparap}(a) and (c)).
This makes the analysis more challenging because the changes in both $\mu_{\text{h}}$ and $\mu_{\text{e}}$ are expected to significantly affect the transport response across the spin-flop-like transition.

\section{Conclusion}
In summary, we investigate the magnetic, magnetoelastic, and magnetotransport properties of the rare-earth-based polar semimetals $R$AuGe ($R=$ Dy, Ho, and Gd).
All the compounds show semimetallic multi-carrier nature in transport properties, which is consistent with the nonmagnetic analogue YAuGe.
DyAuGe, HoAuGe, and GdAuGe exhibit metamagnetic transitions suggesting the frustrated magnetism in the triangular lattice of rare earth ions.
The anomalous properties in GdAuGe reported in Ref. \cite{ram2023multiple} have been reproduced in the self-flux grown crystals, while the detailed differences in magnetism and transport properties suggest that there are qualitative variations depending on the growth conditions.
We identify AHE in DyAuGe and HoAuGe and succeed in quantitatively estimating the anomalous Hall conductivity: $\sim 1200$ S/cm for $R=$ Dy and $\sim 530$ S/cm for $R=$ Ho at $T=1.8$ K in the magnetization saturation regime.
The present study clarifies the magnetotransport properties in $R$AuGe and suggests a connection between electronic and magnetic structures in polar semimetals giving rise to a large AHE.

\section{Acknowledgements}
T.K. was supported by Ministry of Education Culture Sports Science and Technology (MEXT) Leading Initiative for Excellent Young Researchers (JPMXS0320200135) and Inamori Foundation.
This study was supported by Japan Society for the Promotion of Science (JSPS) KAKENHI Grant-in-Aid (No. 21K13874, 20J10988, 23K13068, 22K14010, JP19H05826, 19H01835).
This work was partly performed using the facilities of the Materials Design and Characterization Laboratory in the Institute for Solid State Physics, the University of Tokyo.
The synchrotron radiation experiments were performed at SPring-8 with the approval of the Japan Synchrotron Radiation Research Institute (JASRI) (Proposal No. 2024A1511).
The authors thank A. Nakano and Y. Nakamura for supporting synchrotron x-ray diffraction experiments, L. Ye for fruitful discussion, and A. Ikeda for generously allowing the use of optical sensing instrument (Hyperion si155, LUNA) for thermal expansion and magnetostriction measurements.

\bibliography{reference}

\end{document}